\documentclass[iop]{emulateapj}

\usepackage{apjfonts}
\usepackage{pbox}
\usepackage{bm}
\usepackage{CJK}
\usepackage{color}

\shorttitle{Measuring 14 elemental abundances with LAMOST}
\shortauthors{Ting et al.}

\begin{document}

\begin{CJK*}{UTF8}{gbsn}
\title{Measuring 14 elemental abundances with $R=1$,$800$ LAMOST spectra}
\author{Yuan-Sen Ting (丁源森)\altaffilmark{1,2,3,4,5}, Hans-Walter Rix\altaffilmark{2}, Charlie Conroy\altaffilmark{6}, Anna Y.~Q. Ho\altaffilmark{7}, Jane Lin\altaffilmark{1}}
\altaffiltext{1}{Research School of Astronomy and Astrophysics, Mount Stromlo Observatory, Cotter Road, Weston Creek, ACT 2611, Canberra, Australia}
\altaffiltext{2}{Max Planck Institute for Astronomy, K\"onigstuhl 17, D-69117 Heidelberg, Germany}
\altaffiltext{3}{Institute for Advanced Study, Princeton, NJ 08540, USA}
\altaffiltext{4}{Department of Astrophysical Sciences, Princeton University, Princeton, NJ 08544, USA}
\altaffiltext{5}{Observatories of the Carnegie Institution of Washington, 813 Santa Barbara Street, Pasadena, CA 91101, USA}
\altaffiltext{6}{Harvard--Smithsonian Center for Astrophysics, 60 Garden Street, Cambridge, MA 02138, USA}
\altaffiltext{7}{Cahill Center for Astrophysics, California Institute of Technology, 1200 E. California Boulevard, Pasadena, CA 91125, USA}
\slugcomment{Submitted to ApJL}

%
%
%
%
%
%
\begin{abstract}
The LAMOST survey has acquired low-resolution spectra ($R=1$,$800$) for 5 million stars across the Milky Way, far more than any current stellar survey at a corresponding or higher spectral resolution. It is often assumed that only very few elemental abundances can be measured from such low-resolution spectra, limiting their utility for Galactic archaeology studies. However, \citet{tin17} used {\it ab initio} models to argue that low-resolution spectra should enable precision measurements of many elemental abundances, at least in theory. Here we verify this claim in practice by measuring the relative abundances of 14 elements from LAMOST spectra with a precision of $\lesssim 0.1$ dex for objects with S/N$_{\rm LAMOST}\;\gtrsim30$ (per pixel). We employ a spectral modeling method in which a data-driven model is combined with priors that the model gradient spectra should resemble {\it ab initio} spectral models. This approach assures that the data-driven abundance determinations draw on physically sensible features in the spectrum in their predictions and do not just exploit astrophysical correlations among abundances. Our analysis is constrained to the number of elemental abundances measured in the APOGEE survey, which is the source of the training labels. Obtaining high quality/resolution spectra for a subset of LAMOST stars to measure more elemental abundances as training labels and then applying this method to the full LAMOST catalog will provide a sample with more than 20 elemental abundances that is an order of magnitude larger than current high-resolution surveys, substantially increasing the sample size for Galactic archaeology.
\end{abstract}


%
%
%
%
%
%

\section{Introduction}
\label{sec:introduction}

Galactic archaeology has garnered much momentum in the last few years with the advent of multi-object spectroscopic surveys of stars across our Galaxy, such as APOGEE \citep{maj15,hol15,sds16}, GALAH \citep{des15,mar17}, Gaia-ESO \citep{gil12,ber16}, RAVE \citep{cas17,kun17} and LAMOST \citep{liu17,xia17b}. Galactic archaeology aims at unraveling the chemical and dynamical evolution of the Milky Way, developing it as an archetype for the galactic evolution of spiral galaxies. This goal requires two main components: studying as many stars as possible in the Milky Way and measuring precise stellar properties, in particular, multiple elemental abundances and stellar ages, along with orbits of these stars.

High-resolution spectroscopy ($R> 20$,$000$) is typically thought to be indispensable for robustly measuring individual abundances for many elements. However, high-resolution spectroscopy is usually restricted to the brighter stars in the Milky Way. Low-resolution survey, such as LAMOST, on the other hand, can collect a much larger sample, but so far only a few elemental abundances (C, N, Fe, $\alpha$-enhancement) have been measured from LAMOST spectra \citep{ho17b,ho17a,xia17a}. Recently, \citet{tin17} showed that, at least in theory, low-resolution spectra contain as much spectral information as high-resolution spectra given the same exposure time and CCD pixels and should be able to measure many ($> 10$) elemental abundances. 

Even if detailed abundance information is in principle contained in low-resolution spectra, there are concerns whether it can be extracted in practice: continuum placement and {\it ab initio} model imperfection becomes increasingly problematic at low-resolution. To alleviate these problems, in this Letter we propose an approach that combines data-driven \citep{nes15a,cas16} with {\it ab initio} \citep{tin16,rix16} spectral model fitting. By imposing priors on the data-driven model that are informed by synthetic spectral models, we steer the data-driven approach to pick up the right features for each element at low-resolution. We demonstrate that this method can measure $> 10$ element abundances (to $\lesssim 0.1$~dex) for the $R\simeq 1,800$ LAMOST spectra, opening entirely new opportunities for Galactic archaeology.

This Letter outlines the method and presents a test on a subset of the full LAMOST dataset. We will be exploring the scientific implications of measuring 14 elemental abundances for the whole LAMOST sample ($> 10^6$ stars) in a forthcoming companion paper (Lin et al., in prep.).
\begin{figure*}
\centering
\includegraphics[width=1.0\textwidth]{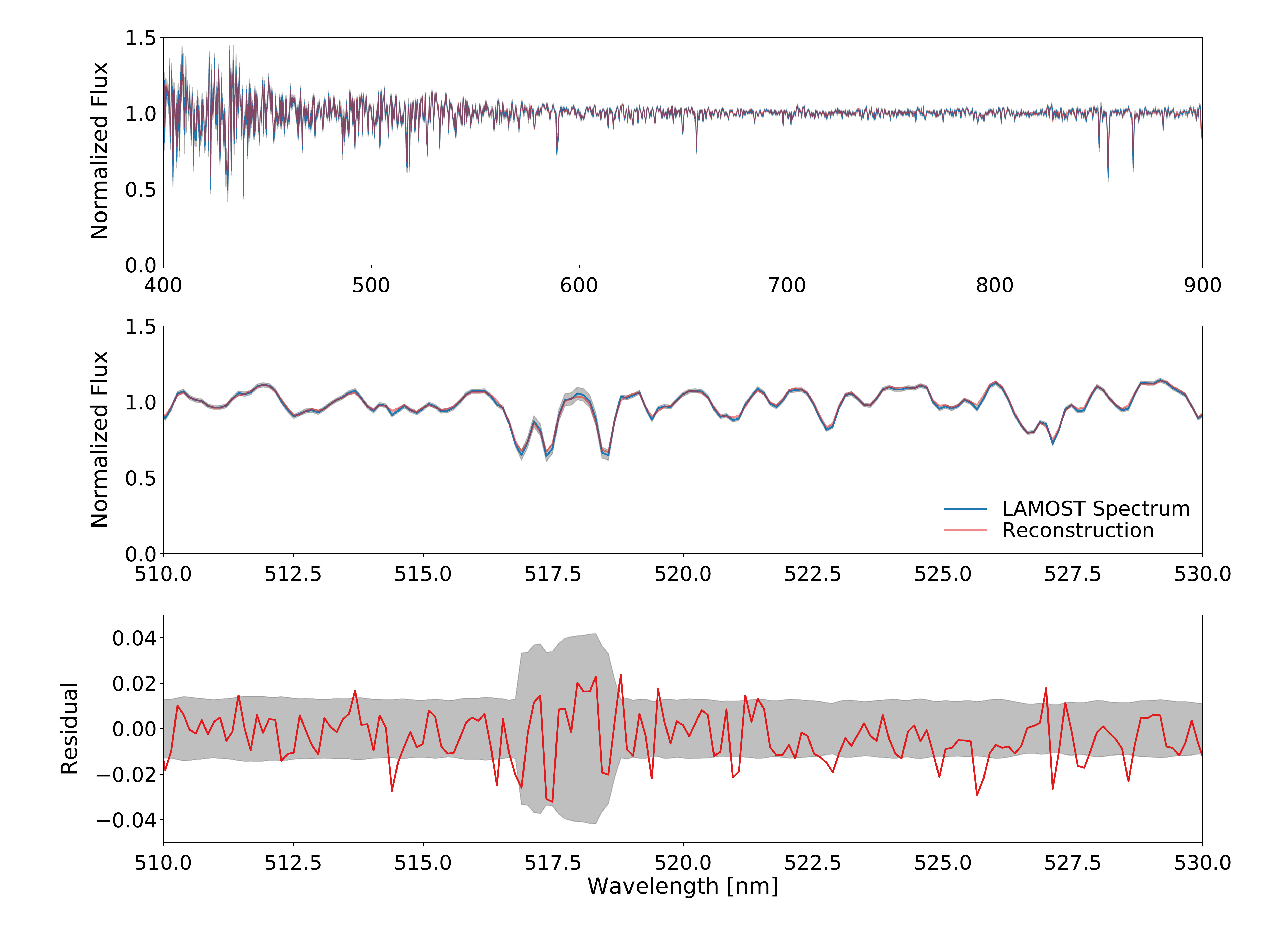}
\caption{Reconstruction of LAMOST spectra using the method presented in this study, which is data-driven but with priors from {\it ab initio} models. In the top and middle panels, the blue lines and the gray shaded regions show a LAMOST observed spectrum and its uncertainties, and the red lines illustrate the model spectrum corresponding to the APOGEE labels for this star. The middle panel is a zoom-in to a portion of the top, with the bottom panel showing the data-model residuals, demonstrating that they are consistent with the observational uncertainties of the spectrum.}
\label{fig1}
\end{figure*}

\begin{figure*}
\centering
\includegraphics[width=1\textwidth]{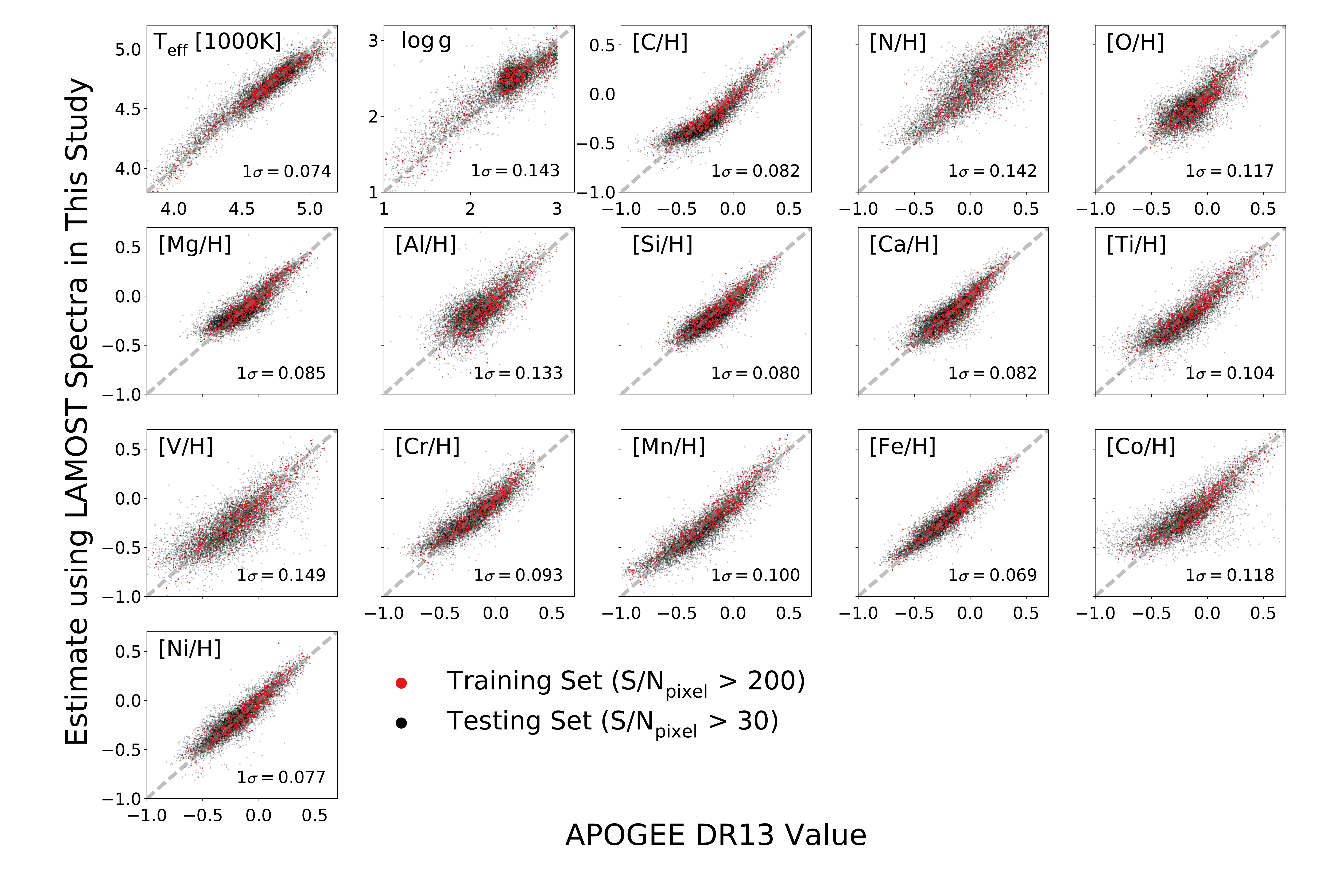}
\caption{Cross validation,  testing the quality of stellar label estimate from LAMOST spectra with the method presented in this study. The $x$-axis shows the APOGEE DR13 values, and the $y$-axis shows our estimates using low-resolution LAMOST spectra for the same stars. The  red points show the leave-none-out test on training data with ${\rm S/N}_{\rm LAMOST} > 200$. The black points show the independent test data with signal-to-noise per pixel of $30 < {\rm S/N}_{\rm LAMOST} < 200$. The $1\sigma$ values at the bottom of each panel show the variance between the APOGEE values to our LAMOST estimates of the {\em independent testing data}. We show that even for the LAMOST spectra with $R=1$,$800$ and $30 < {\rm S/N}_{\rm LAMOST} < 200$, we can recover elemental abundances to a precision of $\sim 0.1\,$dex.}
\label{fig2}
\end{figure*}

\begin{figure*}
\centering
\includegraphics[width=1.0\textwidth]{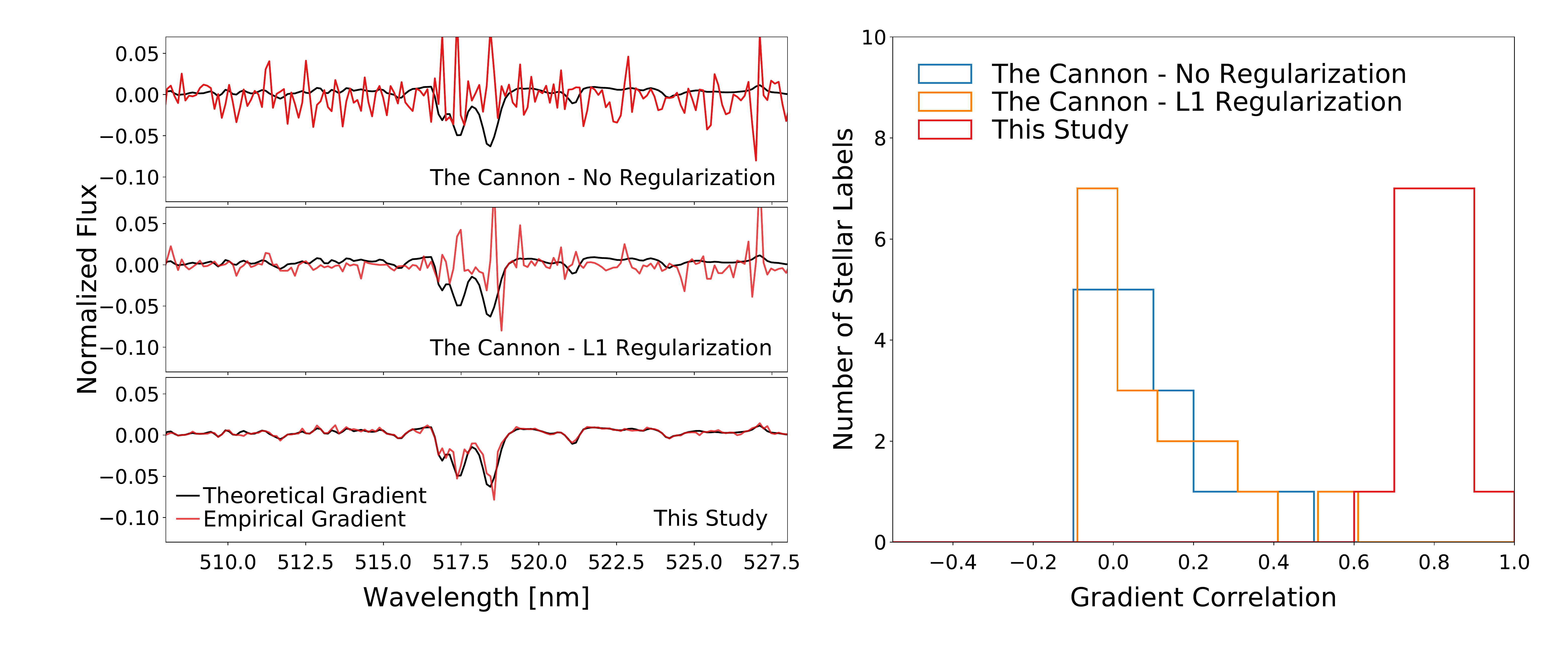}
\caption{Illustration that our spectral modeling method predicts stellar labels from physically sensible spectral features. The left panels show the spectral region near the MgI b triplet. The theoretical gradient spectrum (with respect to [Mg/H]) is in black, and the gradient spectra from purely data-driven models are in red. Both basic data-driven models with (middle panel) or without (top panel) L1 regularization fail to pick up the right feature at $R=1$,$800$, unlike the approach in this study (bottom panel) which is a data-driven model with {\it ab initio} prior. By design, including the theoretical prior makes sure that we draw abundances from physically sensible features. The right panel extends this verification to all 16 stellar labels across the entire LAMOST wavelength range The histograms show the cross-correlation of the model gradient spectra with the theoretical gradient spectra of all 16 stellar labels. Higher cross-correlation value indicates a better agreement between the data-driven model and the theoretical expectation. The gradient spectra from the approach in this study have a greater agreement with the theoretical expectation than the canonical {\it Cannon} approaches.}
\label{fig3}
\end{figure*}

%
%
%
%
%
%

\section{Methods}
\label{sec:approach}

Low-resolution stellar spectra can be fit with {\it ab initio} models, interpolating among a set of synthetic stellar spectra \citep{tin17}. But to derive accurate stellar labels, stellar parameters and element abundances, would then require theoretical model spectra whose systematic shortcomings are negligible; such models do not currently exist \citep[see][]{kur96,kur03,kur05,hau99,gus08,smi14}. This shortcoming has led to the exploration of data-driven approaches, e.g., {\it The Cannon}. Those data-driven models presume that the stellar labels of some observed spectra (the "training set") are known accurately and precisely, which are used to build a pixel-by-pixel model of the spectrum. That same model can then be used to estimate stellar labels for spectra obtained from the same experimental setup not contained within the training set. The advantage of this approach over {\it ab initio} fitting is that, by construction, such data-driven approaches do not suffer from systematic errors of synthetic spectral models (but such methods inherit the biases of the training set). Therefore, they have been very successful in determining {\em precise} (not necessarily accurate) stellar labels from spectra (\citealt{nes15a,cas16,ho17b}).
 
But data-driven models have critical interpretive limitations, in particular in the case where the training spectra are noisy and the stellar labels to be determined are strongly correlated for astrophysical reasons, such as the abundances of iron-peak elements [Fe/H], [Ni/H], [Cr/H], etc. A data-driven model may ``learn" that the data constrain [Fe/H] well and that [Ni/H] and [Fe/H] are astrophysically correlated. It predicts [Ni/H] correctly (based potentially in good part on Fe spectral features) but it might not actually ``measure" this [Ni/H] abundance off the spectrum. Objects with unusual [Ni/Fe] abundances might then be by construction undetectable.

 \citet{cas17} tackled this problem by implementing an L1 regularization on {\it The Cannon}, penalizing models for unneeded non-zero coefficients to prevent data-driven models from over-fitting the data. \citet{cas17} found that optimized L1 regularization for APOGEE spectra and 15 abundances led to spectral model coefficients that were physically plausible. When applying an analogous approach to the LAMOST spectra of far lower resolution, we found it to work well for models with few labels ($\sim 4$), providing model gradient spectra with little aliasing compared to {\it ab initio } models.  However, for 16 labels and the low-resolution LAMOST spectra of interest here we found this not to work well, presumably for two reasons: at higher resolution, the features and their variations are stronger and more prominent, therefore, the noise of the training spectra plays a smaller role, which helps to break this degeneracy. Also with a smaller number of labels, it is easier to find the exact functional form as there are fewer correlated labels for gradient aliasing. In Ting et al. in prep. we also show that a quadratic model is not sufficient to map the flux behavior across a wide parameter range \citep[see also][]{cas17}. Therefore, we will extend and generalize the idea of {\it The Cannon} with two new ingredients.

First, we generalize the label-dependent flux prediction (from a polynomial) to a non-parametric model that will be fully expounded in a forthcoming paper ({\it The Payne}, Ting et al. in prep.). In brief, instead of imposing an explicit quadratic function, we apply neural networks, which look for an approximation function that best describes the variation of flux as a function of stellar labels through a composite of simple ``activation'' functions. The basic idea is to approximate any complex function (spectral flux, as a function of stellar labels) by a composite of simple functions, with the neural net learning the relative weights (scales) and biases (shifts) of the composite function. In this study, we consider a simple neural network architecture that consists of one hidden layer with 100 nodes, using the sigmoid function $1/(1+e^{-x})$ as the activation function. We assume the variation of flux {\em at each wavelength pixel} to be
\begin{equation}
f= w' \cdot g \Bigg[ \sum_{j=1}^{100} \bigg( w_j \cdot g\bigg[\sum_{i=1}^{N_{\rm labels}} \big(w_{j,i} \cdot \ell_i + b_{j,i} \big) \bigg] + b_j \bigg) \Bigg] + b', \nonumber
\label{eq:NN}
\end{equation} 

\noindent
where $g$ is the sigmoid function and $\boldsymbol{\ell}$ is the stellar label. The training step adjusts the weights, $w$, and biases, $b$, minimizing the loss function. The simplest loss function would be minimizing the $\chi^2$ (over all training spectra) making this a (non-polynomial) generalization of {\it The Cannon}. However, this simple loss function might not necessarily favor models that draw the label predicting information from physically sensible and interpretable parts of the spectrum. 

To overcome this issue, the second new ingredient that we implement in this study is assuming a prior on the data-driven model, based on {\it ab initio} spectral models. In essence, we select data-driven models that resemble (but not necessarily equal) the gradient spectra of the {\it ab initio} models. The ``gradient spectra'' in \citet{tin16,rix16}, are defined as the change in the model spectra as we vary each stellar label by a small amount, holding all other labels fixed. Similar to \citet{tin16}, we chose $\Delta T_{\rm eff} = 200\,{\rm K}$, $\Delta \log g = 0.5$ and $\Delta [X/{\rm H}] = 0.2$. We choose a reference point, $\boldsymbol{\ell}_{\rm ref}$ for K-giants ($T_{\rm eff} = 4$,$800\,$K, $\log g= 2.5$ and solar metallicity) since we will fit spectra in this parameter range in this initial study. When working on the full LAMOST catalog which spans a broad range of $T_{\rm eff}-\log g$, one has to adopt a combination of different reference points. How to implement it seamlessly is an aspect that we are exploring for the full catalog paper. With the model prior included, the loss function for one pixel reads\\

\vspace{0.2cm}
\begin{eqnarray}
\mathcal{L}\big(\{f_{\rm obs}\} | \mathbf{w},\mathbf{b}, \{\boldsymbol{\ell}_{\rm obs}\}\big) = \frac{1}{N_S} \sum_{i = 1}^{N_S} \frac{\big(f(\lambda | \boldsymbol{\ell}_{\rm obs,i}) - f_{\rm obs,i}(\lambda)\big)^2}{\sigma_{\rm obs,i}^2(\lambda)} \nonumber \\
+ D_{\rm scale} \cdot \frac{1}{N_{\rm l}} \sum_{j=1}^{N_{\rm l}} \log \frac{|f'(\lambda | \boldsymbol{\ell}_{\rm ref}) - f'_{\rm ab\,initio}(\lambda)|}{|f'_{\rm ab\,initio}|},
\end{eqnarray}

\noindent
where $N_S$ is the number of training spectra, and $N_l$ is the number of labels.

The first term of the loss function is the usual $\chi^2$ minimization. The second term is the prior term taking into account how much the data-driven model gradient spectrum $f'$, convolved to the observed resolving power and $\lambda$-sampling, differs from the corresponding one based on the Kurucz models \citep{kur81,kur93} spectra $f'_{\rm ab\,initio}$ \citep[see appendix of][]{tin17}. This term encapsulates that we have considerable faith in the predictive power of the (quantitatively imperfect) {\it ab initio} models: wavelengths that are deemed (un-)informative by {\it ab initio} models, should also be comparably (un-)informative in the model. For example, if the theoretical gradient at a given wavelength is close to zero, even small deviations of the data-driven gradients from the theoretical gradients will imply severe penalties. On the other hand, if the theoretical gradient at a pixel is strong, $|f'_{\rm ab initio}| \gg 0$, implying that this pixel should be informative about a certain label, then the data can determine the actual model gradient spectra. The $D_{\rm scale}$ is a free hyperparameter to tune the relative importance of the theoretical prior to the fully data-driven approach. With a larger $D_{\rm scale}$, we strengthen the theoretical prior, but we might sacrifice how well we can recover the labels if the theoretical models are not exactly correct. A smaller $D_{\rm scale}$ allows the empirical models to readjust more the strength of each spectral feature, building on the basis of the theoretical models. But if $D_{\rm scale}$ is too small, we will revert to the pure data-driven regime, where the elemental abundances might not draw from physically sensible features. In short, we want the model to look as much like the {\it ab initio} model as possible, without sacrificing the label prediction (as tested by cross-validation). We found that $D_{\rm scale} = 10$ works well in our case -- the cross-validation analysis shows that the precision decreases by a factor of 1.5 compared to the case with $D_{\rm scale} = 0$, but as we will see, the model is more physically plausible with this choice.

%
%
%
%
%
%

\section{Measuring 14 Elemental Abundances from LAMOST Spectra}
\label{sec:results}

We now show how well 16 labels can be determined from LAMOST spectra, $T_{\rm eff}$ $\log g$ and 14 elemental abundances in $[X/{\rm H}]$ (C, N, O, Mg, Al, Si, Ca, Ti, V, Cr, Mn, Fe, Co Ni). We do this with the above spectral model, by transferring this label information in a training step from APOGEE to LAMOST, using $\sim 500$ cross-matched objects between the APOGEE DR13 and LAMOST DR3 catalogs with S/N$_{\rm LAMOST} \;>200$ (per pixel). We only consider giants with $\log g < 3$, as APOGEE DR13 did not derive elemental abundances for dwarfs. There are another $\sim 7500$ overlapping targets with S/N$_{\rm LAMOST} \; >30$, which serve as test and cross-validation spectra. We assume the APOGEE stellar label estimates to be the ground truth, $\boldsymbol{\ell}_{\rm obs}$. We normalize all spectra (LAMOST spectra and Kurucz model spectra) in the same way following \citet{ho17a}, dividing out a version of the spectra that was smoothed with a Gaussian kernel with an FWHM $5\,$nm in width.

We start by illustrating in Fig.~\ref{fig1} how well the model can predict the normalized spectra for LAMOST, given a set of labels from APOGEE. In the top and middle panels, the red lines show the model reconstruction, and the blue lines with the gray band show a LAMOST observed spectrum and its uncertainties. The bottom panel illustrates the residuals of the model compared to the observed spectrum, demonstrating that the model uncertainties are consistent with the observational uncertainties. Fig~\ref{fig2} shows via cross-validation how well we recover stellar labels from LAMOST spectra. The $x$-axis shows the APOGEE DR13 values, and the $y$-axis shows our estimates derived from LAMOST spectra. The red points illustrate the label estimates for the training set of this study and the black points are the independent testing set. The figure demonstrates that even with S/N$_{\rm LAMOST} \;>30$ LAMOST spectra, we can derive elemental abundances that are precise to $0.1$ dex compared to APOGEE estimates. However, a good agreement in this cross validation test alone is not sufficient to confirm that we have measured elemental abundances due to the astrophysical correlations mentioned above, which is what we will verify next.

The left panel of Fig.~\ref{fig3} shows the comparison of the model gradient spectra with the theoretical gradient spectra in three cases focusing on the prominent MgI b triplet. The top panel shows a {\it Cannon} model without regularization; the middle panel includes L1 regularization, and the last panel shows the approach in this study (a data-driven model with {\it ab-initio} prior). For the L1 approach, we adopted a similar approach as in \citet{cas17}, but here we penalize non-zero weights and biases in the neural network. In other words, in the cost function as shown in Eq.~(2), instead of adding a penalty term based on the theoretical prior, we penalize the models with an extra term $\Lambda \sum (|w_i| + |b_i|)$, summing over all weights and biases in the neural net. We tested a wide range of value for $\Lambda$, spanning six orders of magnitude, and chose the $\Lambda$ that gives empirical gradients closest to the theoretical gradients. We also tested the case of a quadratic model, and the results remain qualitatively similar. The figure indicates that at low resolution, the canonical {\it Cannon} approach may predict labels quite well, but does not draw in this prediction from features implied by physical {\it ab initio} models -- simple data-driven models may attribute absorption features to multiple unrelated labels. For instance, Fig \ref{fig3} shows that the Mg label does not pick up all the power of the MgI b triplet. Furthermore, the Mg label picks up features that are not spectrally related to Mg -- we found that some power is attributed to other elemental abundances. On the other hand, the model in this study robustly finds the relevant features of each element because, by design, we require the model to extract elemental abundances information from features predicted by theory. To quantify this point, the right panel of Fig~\ref{fig3} shows the model-to-theoretical correlation of the {\em same} labels across the entire wavelength range for all 16 labels in this study, indicating the extent to which the model picks up the corresponding spectral features. This panel demonstrates that the model in this study draws its label predictions from a much more physically motivated basis than the canonical data-driven approach: it infers abundances from the correct spectral features.

%
%
%
%
%
%

\section{Discussion and Outlook}
\label{sec:discussions}

In this study, we demonstrated with real data that one can measure 14 elemental abundances from low-resolution ($R=1$,$800$) optical spectra. This study opens up many new opportunities for Galactic archaeology. Our approach relies on a spectral model that combines a data-driven technique with physically motivated priors drawn from {\it ab initio} spectral models. One implication of this result is that continuum normalization, which is a highly non-trivial procedure at low spectral resolution, is not a significant obstacle to measuring detailed abundance patterns even in the limit of severely blended absorption lines.

\citet{tin17} predicted that we should be able to measure $> 20$ elemental abundances from LAMOST spectra. Here we only measured 14 elemental abundances. The fact that we have not attained the full potential of low-resolution spectra may be due to several reasons: First, the method proposed here still relies on data-driven models -- we can only measure elemental abundances that have other independent estimates from high-resolution counterparts, in this case, APOGEE. Since APOGEE is an infrared survey with fewer elemental abundances measured, we are limited in the number of labels we can transfer. But we note that the most difficult element we measured, according to the theoretical prediction, is O:\footnote{In a companion paper, we lay out how we determine the abundances of O, though there are no strong O features in the optical at $R=1$,$800$ (Ting et al., in prep.).} it ranks $\sim 20$th in the Cramer-Rao bound calculation (see figure~4 in \citealt{tin17}). Therefore, with other estimates from multi-object optical high-resolution spectrographs (GALAH, GES) soon becoming publicly available, we have every expectation that we can obtain $\sim 20$ elemental abundances for LAMOST with this approach.

Interestingly, although Na is measured in APOGEE and has strong features in LAMOST (the NaI D line), we found that Na is less precisely measured ($\sigma_{[{\rm Na}/{\rm H}]} \simeq 0.2$) in this study compared to weaker elements, and is therefore omitted in this study. The lack of precise measurements is likely because the NaI D line is strongly contaminated by interstellar absorption. 

Although not shown, we also tried to measure weaker elements in LAMOST that have APOGEE estimates, such as K and S. These elements rank about $\sim 35$th in the Cramer Rao bound calculation. Measuring them would indicate that we can measure $> 20$ elemental abundances from LAMOST. However, the results are not conclusive for these elements. Similar to Na, they also have a large spread when compared to the APOGEE estimates. The spread is not surprising because the absorption features from these elements are very shallow in the LAMOST spectra and hence are more susceptible to the uncertainties in the training labels and spectra as well as the errors in the line list and continuum normalization. Restricting the training set to an even higher cutoff, e.g., S/N$_{\rm LAMOST} \;>300$ tentatively suggests that we can measure these elements, but the size of the training set becomes too small to be reliable.

How do our results compare with the theoretical limit? We calculate the Cramer-Rao bound similar to \citet{tin17} but at S/N$_{\rm LAMOST} \;=30$, and with the continuum normalization procedure adopted in this study, we find that we should be able to measure most elemental abundances in this study to a precision of $\sim 0.05$ dex. So we are performing about two times worse than the theoretical limit. Not attaining the absolute theoretical limit is not entirely surprising. For example, we assume that the labels from APOGEE are ground truth when we train the model, which is likely untrue in detail and could compromise the model. Correspondingly, it is also worthwhile to further improve the accuracy of stellar parameters and elemental abundances through, for e.g., 3D non-LTE calculations \citep[e,g.,][]{ama17}.

Finally, while in this study we propose that including theoretical priors can improve how robustly we draw abundance measurements from sensible spectral features. In some regimes a purely data-driven approach works if a suitably tailored regularization and training data are adopted. But in the regime of low-resolution spectra and of many elemental abundances, the gradients of data-driven model tend to show quite severe aliasing, compared to {\it ab initio} models; this may be as the spectra features are shallower, given the same noise in the training spectra, the noise can play a more significant role. Also as the number of correlated labels increases, the problem becomes inherently more degenerate, and it is harder to avoid gradient aliasing among the many labels if the training data are noisy. We cursorily explored that this can be alleviated with (a) more training data, (b) high S/N training spectra and (c) training data that have less-correlated labels. Nonetheless, imposing a prior on the data-driven model gradients to resemble {\it ab initio} models, unless the training data suggest otherwise, seems like a new and effective way forward.

This study has shown that we can deliver many elemental abundances from low-resolution spectra provided that there is sufficient overlap with high-resolution spectra to serve for calibration, making low-resolution surveys excellent and highly complementary tools to the ongoing high-resolution studies.  One implication of this study is that the low and high resolution approach of upcoming Galactic archaeology surveys such WEAVE and 4MOST might prove to be very powerful. Finally, (re)analyzing spectra from the completed/ongoing surveys such as SEGUE and LAMOST as well as the upcoming DESI survey using this method can provide an unprecedented stellar inventory for Galactic archaeology.

%
%
%
%
%

\section{Acknowledgments}
\label{sec:acknoledgements}
YST is supported by the Australian Research Council Discovery Program DP160103747, the Carnegie-Princeton Fellowship and the Martin A. and Helen Chooljian Membership from the Institute for Advanced Study at Princeton. HWR's research contribution is supported by the European Research Council under the European Union's Seventh Framework Programme (FP 7) ERC Grant Agreement n. [321035] and by the DFG's SFB-881 (A3) Program. CC acknowledges support from NASA grant NNX13AI46G, NSF grant AST-1313280, and the Packard Foundation. AYQH is supported by a National Science Foundation Graduate Research Fellowship under Grant No. DGE‐1144469.

%
%
%
%
%
%

\end{CJK*}


\end{document}